\documentclass[fleqn,twoside]{article}
\usepackage{espcrc2}

\usepackage{graphicx}

\newcommand{\AmS}{{\protect\the\textfont2
  A\kern-.1667em\lower.5ex\hbox{M}\kern-.125emS}}

\title{Study of the SU(2) High Density System with Wilson Fermion}

\author{S. Muroya\address{Tokuyama Women's Coll. Tokuyama, 745-8511, Japan},
        A. Nakamura\address[IMC]{IMC, Hiroshima University,  
        Higashi-Hiroshima 739-8521,  Japan }
        and
        C. Nonaka\addressmark[IMC]\thanks{Presenter}
        }
       
\begin{document}

\begin{abstract}
We investigate high density state of SU(2) QCD 
in the case of the Wilson fermions with Iwasaki improved action.
Based on the Woodbery formula, the ratio of fermion determinants is 
evaluated at each step of the Metropolis link update.   At $\beta=0.7$, 
for $\kappa = 0.150$, and $\kappa = 0.175$, 
we calculate thermodynamical quantities, such as,  baryon number density,  Polyakov line, and the energy density of gluon sector with chemical potential $\mu$=0 to 0.9 on the $4^{3} \times 8$ lattice.  The susceptibility of the Polyakov line to the chemical  potential shows peak at around $\mu = 0.7$ which indicates the deconfinement phase transition. Behavior of the meson and diquark propagators with finite chemical potential are also investigated at both side of the peak.
.
\vspace{1pc}
\end{abstract}

\maketitle

\section{Introduction}

High density state of the strongly interacting matter is a very attracting problem 
in the connection to the recent experiments of the Relativistic Heavy Ion 
Collider at BNL \cite{QM}.  Physics of extremely high density state is also 
important for the  understanding the astrophysical objects such as neutron star.  
Usually in the statistical physics, high density state is investigated with use of the 
ground canonical ensemble with chemical potential which is introduced as a Lagrange Multiplier for the conserved charge.  
However, as is well known, chemical potential 
makes action of the QCD complex and causes a problem  in the  numerical convergence  of simulation based on the monte calro method \cite{Simon}.  

SU(2) gauge theory is the simplest non-abelian gauge theory and holds the pseud-real property which keeps action real even with chemical potential.  Though the real QCD is not SU(2) but SU(3),  SU(2) gauge theory is expected to  play an role as a QCD-like theory which informs us important features of the non-abelian field theory \cite{Kogut00}.  SU(2) gauge theory as a residual interaction in the Color Super Conducting state is also one of the promising possibility and the high density state of SU(2) gauge theory can be realized \cite{CCC}.

We calculate thermodynamical  quantities as functions of chemical potential.  Ground canonical ensemble states are generated through update based on the Woodbery formula \cite{Barbour87}. Iwasaki improved action is adopted for the gauge action. Though the improved action seems to stabilize the simulation better than the naive Wilson action \cite{Muroya01},  our simulation  still shows instability at around $\mu =1.0$.    However, the susceptibility of the Polyakov line $\langle \frac{\partial L}{\partial \mu} \rangle$ shows peak at around $\mu = 0.7$ which suggests the deconfinement phase transition.  Therefore, our simulation can correspond to the state  around across the phase transition.  

\section{Thermodynamical quantities}

With  use of  Wilson fermions and Iwasaki improved action we calculate thermodynamical quantities, such as Polyakov line, gluon energy density and baryon number density with $4^{3} \times 8$ lattice.  
The expectation value of the baryon number density is given by,
\begin{equation}
<n> = \frac{1}{\beta V_s} \frac{\partial}{\partial\mu} \log Z
\end{equation} \noindent
where $V_s$ is the spatial volume $N_{x} \times N_{y} \times N_{z}$.
Energy density, $\varepsilon$, is given as, 
\begin{equation}
\varepsilon = \frac{1}{V_s}\left(- \frac{\partial}{\partial \beta} 
+ \frac{\mu}{\beta} \frac{\partial}{\partial \mu} \right) \log Z
\end{equation}
The derivative of the partition function is composed of two parts,
\begin{equation}
(\log Z)'= \frac{1}{Z} \int {\cal D}U{\cal D}{\bar \psi}{\cal D}{\psi}
(-S'_{G}-S'_{F})e^{(-S_{G}-S_{F})},
\end{equation}
where $S_{G}$ and $S_{F}$ are gluon action and fermion action, respectively.
We denote the contribution of the gluon action part (first term in the r.h.s. of eq.(3)) by  gluon energy density.
Figure 1 displays  Polyakov line, gluon energy density and baryon number density 
as functions of $\mu$ where $\beta = 0.7$ and $\kappa = 0.150$, respectively.  
All quantities start to have non-zero value at about $\mu=0.4$ and rise up with
chemical potential in the region $0.4< \mu < 0.8$.
None of them is order parameter of the phase transition in the exact sense, and
since the lattice size is small, no sharp change is seen. 
However,  growing up  of the these quantities  indicates that quarks and gluons
become free from the confinement force at finite chemical potential about $0.4< \mu < 0.8$.

%
\begin{figure}
\begin{center}
\includegraphics[width=.9 \linewidth]{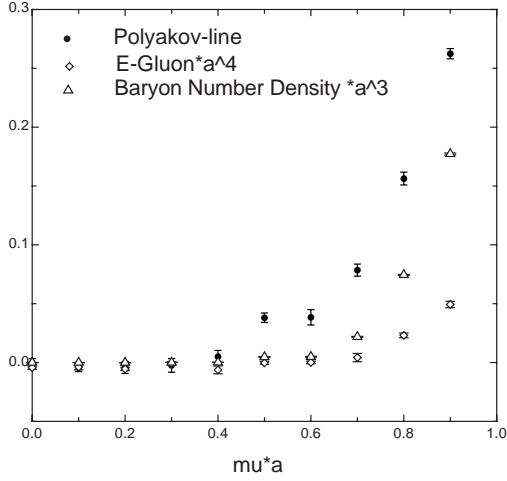}
\end{center}
\caption{Polyakov line, gluon energy density, and baryon number density as a function of  chemical potential $\mu$.}
\label{Fig-density}
\end{figure}

%
\begin{figure}
\begin{center}
\includegraphics[width=.9 \linewidth]{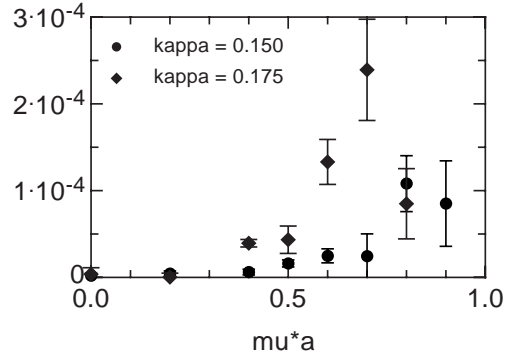}
\end{center}
\caption{Susceptibility of the Polyakov line as a function of  chemical Potential $\mu$. The circle plots and diamond plots correspond to the $\kappa = 0.150$ and $\kappa = 0.175$, respectively.  The peak point seems to  depend on the hopping parameter $\kappa$}
\label{Fig-sus}
\end{figure}

Figure 2 shows susceptibility of the Polyakov line as the functions of 
the chemical potential $\mu$.  The peak appears at around $\mu=0.7$ which suggests 
that our simulation crosses the deconfinement phase transition. However, peak point seems to depend on the hopping parameter $\kappa$.  Larger hopping parameter seems to give lower phase transition $\mu_{C}$.

\section{Propagators}

Figure 3 shows our preliminary results of the propagator of mesons and diquarks with $\kappa = 0.175$.  In the lower figure, at $\mu = 0.0$, pion and scalar diquark(b5b), and a0 and pseudscalar diquark(b1b) are degenerate  each other, as expected.  However, such degeneration of meson and diquark disappears with finite chemical potential.   Because diquark possesses definite baryon number, the effect of the chemical potential on diquark and anti-diquark are in the opposite sign.  Therefore, the symmetry of the propagator in time direction and anti-time direction is broken and asymmetry in nt appears(b1b and b5b in the lower figure of Fig.3). On the other hand, mesons, which have vanishing baryon number, keep symmetric property in nt direction \cite{RCNP}. According to figure 2, $\mu = 0.8$ with $\kappa = 0.175$ is in the region upper side than the peak point.  

\begin{figure}
\begin{center}
\includegraphics[width=.9\linewidth]{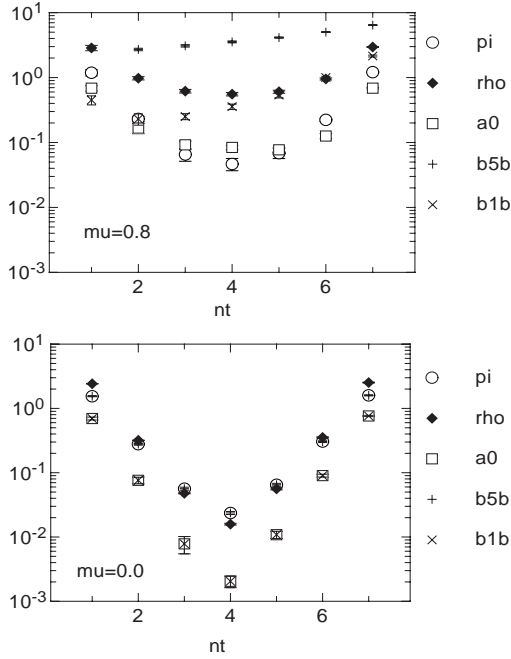}
\end{center}
\caption{Preliminary results of the propagator of the meson and diquark with $\kappa = 0.175$.  The upper figure and 
lower figure stands for the $\mu = 0.8$ and $\mu = 0.0$, respectively.
If we look upon the peak in figure 2 as the deconfinement phase transition point,  
those figures correspond to the mesons diquarks in the state over and under the phase transition}
\label{Fig-Propa}
\end{figure}

\section{Concluding Remarks}

SU(2) finite density state is investigated with Wilson fermion and Iwasaki 
improved action.  Peak of the susceptibility appears at around $\mu = 0.7$, but it seems 
to depend on the $\kappa$.  It means physical size of the lattice spacing will change with $\kappa$ and careful treatment will be required in the procedure of the chiral limit.  The propagators of the mesons and diquark state (Baryon in SU(2)) have also discussed \cite{RCNP}. Detailed analyses for both meson and diquark will appear somewhere else.

\vspace{.2cm}
\noindent
{\bf Acknowledgment}

One of the authors(C.N.) would like to
acknowledge the financial support by the Soryushi Shogakukai.
This work is supported by Grant-in-Aide for Scientific Research by
Monbu-Kagaku-sho, Japan (No.11440080 and No. 12554008).
Simulations were performed on SR8000 at IMC, Hiroshima
University, SX5 at RCNP, Osaka university, SR8000 at KEK and
VPP5000 at Science Information Processing Center, Tsukuba university.

\end{document}